\documentclass[11pt,a4paper]{article}
\usepackage{jcappub}

\usepackage{bm}
\usepackage{epsfig}
\usepackage{graphicx}
\usepackage{amsmath}
\usepackage{amssymb}
\usepackage{amsbsy}
\usepackage{color}
\usepackage{subfigure}
\usepackage{slashed}
\usepackage{afterpage}
\usepackage{psfrag}
\usepackage{hyperref}

\renewcommand\({\left(}
\renewcommand\){\right)}
\renewcommand\[{\left[}
\renewcommand\]{\right]}

\renewcommand\({\left(}
\renewcommand\){\right)}
\renewcommand\[{\left[}
\renewcommand\]{\right]}

\newcommand{\vv}[2]{ \left( \begin{array}{cc}   #1 \\ #2 \end{array} \right)}

\newcommand{\exclude}[1]{}

\begin{document}
\subheader{\hfill CERN-TH-2017-078\\
\hbox to\textwidth{\hfill MPP-2017-85}}

\title{Axion-photon conversion caused by dielectric interfaces:\\
quantum field calculation}

\author[a,b]{Ara N.~Ioannisian,}
\author[b]{Narine Kazarian,}
\author[c]{Alexander J.~Millar}
\author[c]{and Georg G.~Raffelt}

\affiliation[a]{Yerevan Physics Institute, Alikhanian Br.~2, 375036 Yerevan, Armenia}

\affiliation[b]{Institute for Theoretical Physics and Modeling, 375036 Yerevan, Armenia}

\affiliation[c]{Max-Planck-Institut f\"ur Physik (Werner-Heisenberg-Institut),\\
F\"ohringer Ring 6, 80805 M\"unchen, Germany}

\emailAdd{ara.ioannisyan@cern.ch}
\emailAdd{narinkaz@gmail.com}
\emailAdd{millar@mpp.mpg.de}
\emailAdd{raffelt@mpp.mpg.de}

\abstract{Axion-photon conversion at dielectric interfaces, immersed in a
near-homogeneous magnetic field, is the basis for the dielectric haloscope
method to search for axion dark matter. In analogy to transition radiation, this
process is possible because the photon wave function
is modified by the dielectric layers (``Garibian wave function'') and
is no longer an eigenstate of momentum. A conventional first-order
perturbative calculation of the transition probability between a
quantized axion state and these distorted photon states provides the microwave
production rate. It agrees with previous results based
on solving the classical Maxwell equations for the combined system of
axions and electromagnetic fields. We argue that in general the average
photon production rate is given by our result, independently
of the detailed quantum state of the axion field.
Moreover, our result provides a new perspective on axion-photon conversion in
dielectric haloscopes because the rate is based on an overlap integral
between unperturbed axion and photon wave functions, in analogy to the
usual treatment of microwave-cavity haloscopes.
}

\maketitle

\clearpage

\section{Introduction}
\label{sec:intro}

\begin{figure}[ht]
\centering
\includegraphics[width=4cm]{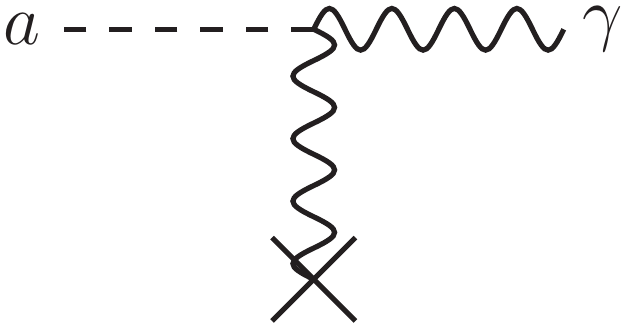}
\caption{Primakoff conversion between axions and photons.}
\label{fig:primakoff}
\end{figure}

The generic two-photon interaction of axions and axion-like particles
provides the basis for many possible astrophysical implications and
experimental search strategies for these elusive particles. The
interaction Lagrangian is usually written in the form
\begin{equation}\label{eq:Lagam}
{\cal L}_{a\gamma}=g_{a\gamma}\,{\bf E}\cdot{\bf B}\,a\,,
\end{equation}
where ${\bf E}$ and ${\bf B}$ are the electric and magnetic fields,
$a$ the axion field, and $g_{a\gamma}$ a coupling constant with
dimension energy$^{-1}$. One obvious consequence, the decay
$a\to\gamma\gamma$, is strongly phase-space suppressed for the very
low-mass axions that are usually considered. Therefore,
instead one frequently considers axion-photon conversion $a\to\gamma$
in the presence of external sources of electric or magnetic fields,
i.e., one of the two photons is virtual and usually static. The
simplest manifestation is Primakoff conversion
(figure~\ref{fig:primakoff}), where an axion converts to a photon (or
the other way round) in the Coulomb field of a charged particle
\cite{Dicus:1978fp}.  This process leads to axion emission in hot
stellar plasmas, notably in the Sun \cite{Raffelt:1985nk}.

A somewhat different kinematical situation arises when axions or
photons propagate in a macroscopic near-homogeneous magnetic field,
for example solar axions propagating in the bores of the CAST dipole
magnet directed toward the Sun \cite{Anastassopoulos:2017ftl}.  ``Primakoff
scattering'' is now in the forward direction and the primary axion and
secondary photon remain coherent with each other. Back-conversion
effects can be important, depending on detailed parameters. In this
case, an intuitive description is in terms of axion-photon
oscillations, analogous to neutrino flavor oscillations, here enabled
by the external transverse $B$-field that mixes axions with photons
\cite{Raffelt:1987im}.  This situation can also be seen as Primakoff
scattering where the formation zone of the secondary photon is the
entire length of the CAST magnet, or even astronomical distances in
the case of photon-axion conversion in astrophysical $B$-fields.

The axion-photon oscillation picture encompasses both a description in
terms of classical or quantum fields. In the presence of a strong
external classical magnetic field ${\bf B}_{\rm e}$, the effective
interaction $g_{a\gamma}\,{\bf E}\cdot{\bf B}_{\rm e}\,a$ between the
photon and axion radiation fields is bilinear so that this ``flavor
oscillation'' can be interpreted as a Bogoliubov transformation
between the axion and photon creation or annihilation operators
\cite{Raffelt:1991ck}. In other words, the same formalism applies if
the initial beam consists of individual axions or photons or of a
highly-occupied and essentially classical laser beam. The outgoing
radiation inherits its quantum properties from the initial beam.

We here consider a yet different kinematical situation motivated by
the usual haloscope method to search for axion dark matter
\cite{Sikivie:1983ip}.  The favored dark matter axion mass is broadly
in the $\mu$eV range. The small galactic virial velocity of around
$10^{-3}\,c$ thus implies a de~Broglie wavelength much larger than the
experimental apparatus. It consists of a strong laboratory magnet to
couple the dark matter axion field with the electromagnetic (EM) field and
to create microwave radiation at a detectable level.  The traditional
cavity experiments use a microwave resonator, i.e., through
${\bf B}_{\rm e}$ the axion field drives EM eigenmodes of the
cavity. This process is particularly effective when the axion mass
$m_a$ matches a cavity resonance.
The traditional calculation of microwave
production uses the equations of motion for the
classical axion field, taken to be essentially homogeneous,
and for the classical cavity mode, taken to be
damped by friction and by the detector which
absorbs power. A perturbative calculation provides the power output in
terms of an overlap integral
$g_{a\gamma} a\int d^3{\bf x}\,{\bf B}_{\rm e}({\bf x})\cdot {\bf E}_{\rm cav}({\bf x})$,
where ${\bf E}_{\rm cav}({\bf x})$ is the electric field configuration of the
used resonant cavity mode.

A new variation of this scheme is the dielectric haloscope approach,
where the conversion of dark matter axions to microwave photons is enabled by
immersing one or more dielectric layers in a strong laboratory field
\cite{Horns:2012jf,Jaeckel:2013sqa,Jaeckel:2013eha,TheMADMAXWorkingGroup:2016hpc,Millar:2016cjp}.
The main motivation is the difficulty of building a large volume of
resonant cavities at relatively large axion masses. The reference
value is $m_a=100~\mu{\rm eV}$, corresponding to a microwave frequency
of 25~GHz, wavelength of 12~mm, and axion de~Broglie wave length of
12~m. The transition between dark matter axions and free microwaves is
suppressed by energy-momentum conservation unless ${\bf B}_{\rm e}$
has a strong spatial variation to provide the required momentum
transfer. In the dielectric haloscope approach, instead the photon
wave function is distorted by the dielectric layers to achieve the
same effect. One way to calculate the emitted microwave power is to
solve Maxwell's equations, augmented with the axion terms, in the
presence of dielectrics \cite{Millar:2016cjp}. Essentially one
determines the exact wave functions of the combined axion-photon
system in the presence of ${\bf B}_{\rm e}$ and the dielectric layers,
matching at every dielectric interface the
axion-induced electric field with those of transmitted and reflected
EM waves.

We here calculate the produced microwave power in a somewhat
alternative way, motivated by the traditional treatment of transition
radiation \cite{Ginzburg:1979wi,Ginzburg:1990}.
A charged particle, moving uniformly on a straight line,
does not radiate due to energy-momentum conservation. However, this is
different in the presence of dielectrics because free photons are now
solutions of Maxwell's equations in the presence of dielectrics, i.e.,
they are no longer eigenstates of momentum. Likewise, for a dielectric
haloscope we can perform a conventional first-order perturbative
calculation for the
transition between dark matter axions and these distorted photons. The
result will be given in terms of a matrix element which effectively
is an overlap
integral similar to the traditional cavity haloscope calculation, but
now involving photon ``scattering states''
rather than the ``bound state'' excitations of a cavity.

This approach gives a new perspective on the conversion of axions to photons inside a dielectric haloscope, and clarifies the relationship between the overlap integral and classical calculations. While both are mathematically equivalent, they evoke very different physical pictures. While previously developed from the classical equations through cumbersome mathematical transformations in reference~\cite{Millar:2016cjp}, the overlap integral formalism emerges here very naturally from the quantum-field calculation. Further, in the classical treatment many features of the overlap integral formalism, such as the normalisation and choice of the integrated $E$ field, emerged seemingly from nowhere. In the perturbative quantum-field
calculation the integrand will be obvious: one must integrate over the free
Garibian photon wave functions~\cite{Garibyan:1961,Garibyan}, which are uniquely determined up to time-reversals. Thus we gain a deeper understanding of the underlying physics of dielectric haloscopes.


\section{General transition rate}
\label{sec:rate}

\subsection{First-order perturbative transition probability}

We consider a generic situation of a plane-wave axion interacting with
a configuration of external classical static electric or magnetic fields
by virtue of the interaction given in equation~(\ref{eq:Lagam}).
We ask for the decay rate (inverse lifetime) $\Gamma_{a\to\gamma}$
of an axion to convert into a photon.
This could be a plane wave with momentum $\bf k$
or another type of propagating state, notably of the type
caused by the presence of dielectrics. In this case ${\bf k}$ represents
some suitable set of quantum numbers describing the wave function.
Generally we consider situations where translational
invariance is broken by external agents. Therefore, the simplest
approach is to use non-covariant perturbation theory
in the laboratory frame.
The inverse lifetime of a single
quantized axion with energy $\omega_a$ to convert into a single photon
following from elementary time-dependent perturbation theory is
\begin{equation}\label{eq:decayrate}
\Gamma_{a\to\gamma}=2\pi\sum_{\bf k}|{\cal M}|^2\,\delta(\omega_a-\omega_{\bf k})\,.
\end{equation}
Here ${\cal M}=\langle {\rm f}|H_{a\gamma}|{\rm i}\rangle$ is the
non-covariant matrix element
of the interaction Hamiltonian between the initial and final state
and as such has the dimension of energy. We use natural units with $\hbar=c=1$.
If ${\cal M}$ does not depend on ${\bf k}$, equation~\eqref{eq:decayrate}
is Fermi's Golden Rule in its simplest form
$\Gamma=2\pi|{\cal M}|^2\,dN/d\omega$ where
$dN/d\omega$ is the density of continuum final states per unit energy.

To recall the normalisation of the fields
quantized in some large but finite volume $V$
we mention that the axion field has the form
\begin{equation}
\phi=\sum_{\bf p}\frac{1}{\sqrt{2\omega_{\bf p} V}}
\(a_{\bf p} e^{-i(\omega_{\bf p} t-{\bf p}\cdot{\bf r})}
+a^\dagger_{\bf p} e^{i(\omega_{\bf p} t-{\bf p}\cdot{\bf r})}\)\,,\label{eq:axionwavefunction}
\end{equation}
where $a_{\bf p}$ and $a^\dagger_{\bf p}$ are the usual destruction
and creation operators for a quantum of momentum ${\bf p}$ and
energy $\omega_{\bf p}=\sqrt{{\bf p}^2+m_a^2}$.
(Notice that $a_{\bf p}$ and $a^\dagger_{\bf p}$ are
dimensionless and $\phi$ had dimension energy as it should.)
In this case we say that the axion plane waves have amplitude~1.
A similar expression pertains to the propagating polarization components of
the quantized photon field $\cal A$.
For simplicity, we will often separate the normalisation from the photon wave function,
writing ${\cal A}_{\bf k}\equiv A_{\bf k}/\sqrt{2\omega_{\bf k} V}$, where plane-wave
photons correspond to $|A_{\bf k}|=1$. However, our main interest is in photon
wave functions modified by dielectrics so that ${\bf k}$ is a more general
set of quantum numbers and $A_{\bf k}$ is a nontrivial wave function that breaks
translational invariance.

As a next step we assume that the external field is given as ${\bf B}_{\rm e}({\bf r})$
or as ${\bf E}_{\rm e}({\bf r})$ and we consider the transition of
one quantum of the axion field~\eqref{eq:axionwavefunction} to a
photon with quantum numbers ${\bf k}$.
The matrix element of the interaction Hamiltonian between
initial and final quantum states is then found to be
\begin{equation}\label{eq:general-matrix-element}
{\cal M}=\frac{g_{a\gamma}}{2\omega V}\int d^3{\bf r}\,
e^{i{\bf p}\cdot{\bf r}}\,{\bf B}_{\rm e}({\bf r})\cdot{\bf E}^*_{\bf k}({\bf r})
\quad\hbox{or}\quad
{\cal M}=\frac{g_{a\gamma}}{2\omega V}\int d^3{\bf r}\,
e^{i{\bf p}\cdot{\bf r}}\,{\bf E}_{\rm e}({\bf r})\cdot{\bf B}^*_{\bf k}({\bf r})\,,
\end{equation}
where $\omega=\omega_a=\omega_{\bf k}$. Moreover,
${\bf E}_{\bf k}({\bf r})$ or ${\bf B}_{\bf k}({\bf r})$ are the
electric or magnetic field configuration associated with the unnormalised photon wave function $A_{\bf k}$. To check the dimensions of this expression, notice that the external
electric or magnetic fields, in natural units, have dimension (energy)$^2$, $g_{a\gamma}$ has dimension (energy)$^{-1}$, and the amplitudes ${\bf E}_{\bf k}$ or ${\bf B}_{\bf k}$ have dimension (energy) because $A_{\bf k}$ is dimensionless and $E_{\bf k}\sim \omega A_{\bf k}$. Overall ${\cal M}$ therefore has dimension (energy) as it should.

The matrix element equation~\eqref{eq:general-matrix-element} is an overlap integral of
the external EM field, sandwiched between the spatial axion and photon wave functions.
In the haloscope context, the axion momentum will be taken to be vanishingly small. In this sense, the microwave production rate is proportional to an overlap integral of the external EM field configuration with the photon wave function.

\subsection{Primakoff transition}

To connect to a familiar case, we consider Primakoff transition of axions
to photons in the Coulomb field of a charge $Ze$. It represents a heavy
nucleus which does not recoil, so its electric field
${\bf E}_{\rm e}({\bf r})=Ze\,{\bf r}/r^3$ plays the role of an external
static field. The photons are taken to be plane waves with
wave vector ${\bf k}$ so that the final-state photon's magnetic field is
${\bf B}_{\bf k}({\bf r})=i({\bf k}\times{\bm \epsilon})\,e^{i{\bf k}\cdot{\bf r}}$,
where ${\bm\epsilon}$ is a polarization vector. Therefore, the matrix element is
\begin{equation}
{\cal M}=i\,\frac{g_{a\gamma}Ze}{2\omega V}\,({\bf k}\times{\bm \epsilon})\cdot
\int d^3{\bf r}\,\frac{{\bf r}}{r^3}\,e^{-i{\bf q}\cdot{\bf r}}\,,
\end{equation}
where ${\bf q}={\bf k}-{\bf p}$ and the integral is $-i{\bf q}/{\bf q}^2$.
Because $({\bf k}\times{\bm \epsilon})\cdot{\bf k}=0$ we find
$({\bf k}\times{\bm \epsilon})\cdot{\bf q}=({\bf k}\times{\bm \epsilon})\cdot{\bf p}=
({\bf p}\times{\bf k})\cdot{\bm \epsilon}$, i.e.,
\begin{equation}
{\cal M}=\frac{g_{a\gamma}Ze}{2\omega V}\,
\frac{({\bf p}\times{\bf k})\cdot{\bm \epsilon}}{|{\bf p}-{\bf k}|^2}\,.
\end{equation}
The final-state photons could have any polarization, so in the
squared matrix element we must perform a spin sum and find
\begin{equation}
\sum_{\bm\epsilon}|{\cal M}|^2=\(\frac{g_{a\gamma}Ze}{2\omega V}\)^2\,
\frac{|{\bf p}\times{\bf k}|^2}{|{\bf p}-{\bf k}|^4}\,.
\end{equation}
Notice that in
$\sum_{\bf \epsilon} ({\bf p}\times{\bf k})_i{\bm \epsilon}_i({\bf p}\times{\bf k})_j{\bm \epsilon}_j$
we can use $\sum_{\bm \epsilon} {\bm \epsilon}_i{\bm \epsilon}_j=\delta_{ij}$ because
the longitudinal part of ${\bm \epsilon}$ vanishes in $({\bf p}\times{\bf k})\cdot{\bm \epsilon}$.

For the sum over final-state photons we instead integrate using
the density of states $V/(2\pi)^3$ in momentum space and include energy conservation, i.e.,
we use $\sum_{\bf k}\,\delta(\omega_a-\omega_{\bf k})
\to V\omega^2 \int d\Omega_{\bf k}/(2\pi)^3$ with
$\omega=\omega_{\bf k}=|{\bf k}|$. Overall we thus find
\begin{equation}
\Gamma_{a\to\gamma}=\frac{1}{V}\(\frac{g_{a\gamma}Ze}{4\pi}\)^2
\int d\Omega_{\bf k}\,\frac{|{\bf p}\times{\bf k}|^2}{|{\bf p}-{\bf k}|^4}\,,
\end{equation}
in agreement with the literature \cite{Raffelt:1987np}.
For massless axions the rate diverges in the forward direction,
although this effect would moderated in a stellar plasma by screening effects.
For axions at rest (${\bf p}=0$) the rate vanishes because the problem is
now spherically symmetric and
there is no induced oscillating electric or magnetic multipole
that could radiate a photon.

The conversion rate on charged particles would be relevant in a stellar
plasma where the relevant quantity is our result times the number of
nuclei, i.e., instead of $1/V$ it is proportional to the number
density of nuclei.  In other words, the factor $1/V$ represents one
charged target particle in our normalisation volume.

\newpage

\subsection{Helioscope}
\label{sec:helioscope}

As another example we consider the axion helioscope \cite{Sikivie:1983ip},
i.e., a dipole magnet oriented toward the Sun such as
the CAST experiment \cite{Anastassopoulos:2017ftl}. The external magnetic
field is taken to be a constant value ${\bf B}_{\rm e}$, whereas the
final-state photon is taken to be a plane wave so that
${\bf E}_{\bf k}({\bf r})=i\omega_{\bf k}\,{\bm \epsilon}\,e^{i{\bf k}\cdot{\bf r}}$.
Therefore, the matrix element is
\begin{equation}
{\cal M}=i\,\frac{g_{a\gamma}\,{\bm\epsilon}\cdot{\bf B}_{\rm e}}{2V}
\int d^3{\bf r}\,e^{-i{\bf q}\cdot{\bf r}}\,.
\end{equation}
We assume ${\bf B}_{\rm e}$ to be transverse to the axion momentum
${\bf p}$, so only photons with polarization parallel to ${\bf B}_{\rm e}$
are produced. Moreover, our setup has translational invariance in the
$y$-$z$-directions, assuming the axion momentum is in the $x$-direction.
Therefore, the $y$ and $z$ components of momentum are conserved and the
photon momentum must also be along the $x$-direction. Our large quantization
volume is $V=S L$ with $S$ some large area in the $y$-$z$-plane and $L$ some
large distance in the $x$-direction.
The magnetic region itself is taken to have length $\ell$ in the $x$-direction,
so with $\int dy dz =S$ overall the matrix element is
\begin{equation}\label{eq:matrix-helioscope}
  {\cal M}=\frac{g_{a\gamma}B_{\rm e}}{2L}\int_{-\ell/2}^{+\ell/2}dx\,e^{-iq x}
  =\frac{g_{a\gamma}B_{\rm e}}{2L}\,\frac{2\sin(q\ell/2)}{q}\,,
\end{equation}
where $q=k_x-p_x$. Note that there are two possibilities for $q$, depending on whether the photon is emitted forwards or backwards, $  q_\pm=\pm \omega - \sqrt{\omega^2-m_a^2}$, respectively. Energy conservation implies $|k_x|=\omega$.

To sum over final states, the symmetry of our setup dictates that only
photon momenta ${\bf k}$ in the $x$-direction appear, i.e.,
$\sum_{\bf k}\delta(\omega_a-\omega_{\bf k})\to (L/2\pi)\,\int d |k_x|
 \delta(\omega_a-\omega_{k_x})
=(L/2\pi)$ with $|k_x|=\omega_{k_x}$.
We find for the axion transition rate
\begin{equation}
  \Gamma_{a\to\gamma}=\frac{1}{L}\,P_{a\to\gamma}
  \quad\hbox{where}\quad
  P_{a\to\gamma}=\sum_{q=q_\pm}\(g_{a\gamma}B_{\rm e}\,\frac{\sin(q\ell/2)}{q}\)^2\,,
\end{equation}
where the sum is over the two cases of momentum transfer.

For solar axions, the x-ray energies are very large compared with
the Fourier components of the magnetic field region, so we
neglect the fast-oscillating part from $q_{-}=-\omega - \sqrt{\omega^2-m_a^2}$.
In other words, ``wrong direction'' photons are hardly emitted because
the $B$-field region does not end abruptly enough to provide
the required large momentum transfer. On the other hand, the difference term
$q_{+}=\omega - \sqrt{\omega^2-m_a^2}$ can be so small that
$q_{+}\ell\ll 1$ and the conversion rate is then the usual
expression $P_{a\to\gamma}=(g_{a\gamma} B_{\rm e}\ell/2)^2$.
Of course, if we neglect the fast-oscillating part, we are back
to the usual picture of axion-photon oscillations in the spirit
of neutrino flavor oscillations along the magnet pipe.

To interpret the normalisation of our result we notice that
by construction we began with one axion in our normalisation volume
$V=S L$. We can express our result as a photon flux per
unit area and unit time emerging at the ends of the magnetic field
region. Therefore, we divide $\Gamma_{a\to\gamma}$ by the
area $S$, implying that the result is proportional to $1/V$, the number
of axions per normalisation volume. So the emerging photon flux
is proportional to the axion number density $n_a$ as it should be.

In the limit ${\bf p}\to 0$ the helioscope becomes a haloscope,
i.e., we can interpret the result as a photon flux from dark
matter axion conversion emerging in two directions from
a homogeneous magnetic field region (length $\ell$)
which ends abruptly on both sides.
The number density of axions is $n_a=\rho_a/m_a$ with
$\rho_a$ the local dark matter axion mass density. Thus the
number of microwave photons emerging from one side of the
magnetic field region per unit time and unit area is
\begin{equation}
\Phi_\gamma=\frac{\rho_a}{m_a}\,
\(g_{a\gamma}B_{\rm e}\,\frac{\sin(m_a\ell/2)}{m_a}\)^2\,.
\end{equation}
In contrast to the Primakoff process on a single charge, we
here obtain a nonvanishing photon production rate in the limit of
vanishing axion velocity. The external magnetic field defines
a vector to produce dipole radiation driven by the oscillating
homogeneous axion field.


\section{Dielectric haloscope}
\label{sec:haloscope}

\subsection{Haloscope setup}

\begin{figure}[ht]
\centering
\includegraphics[width=6cm]{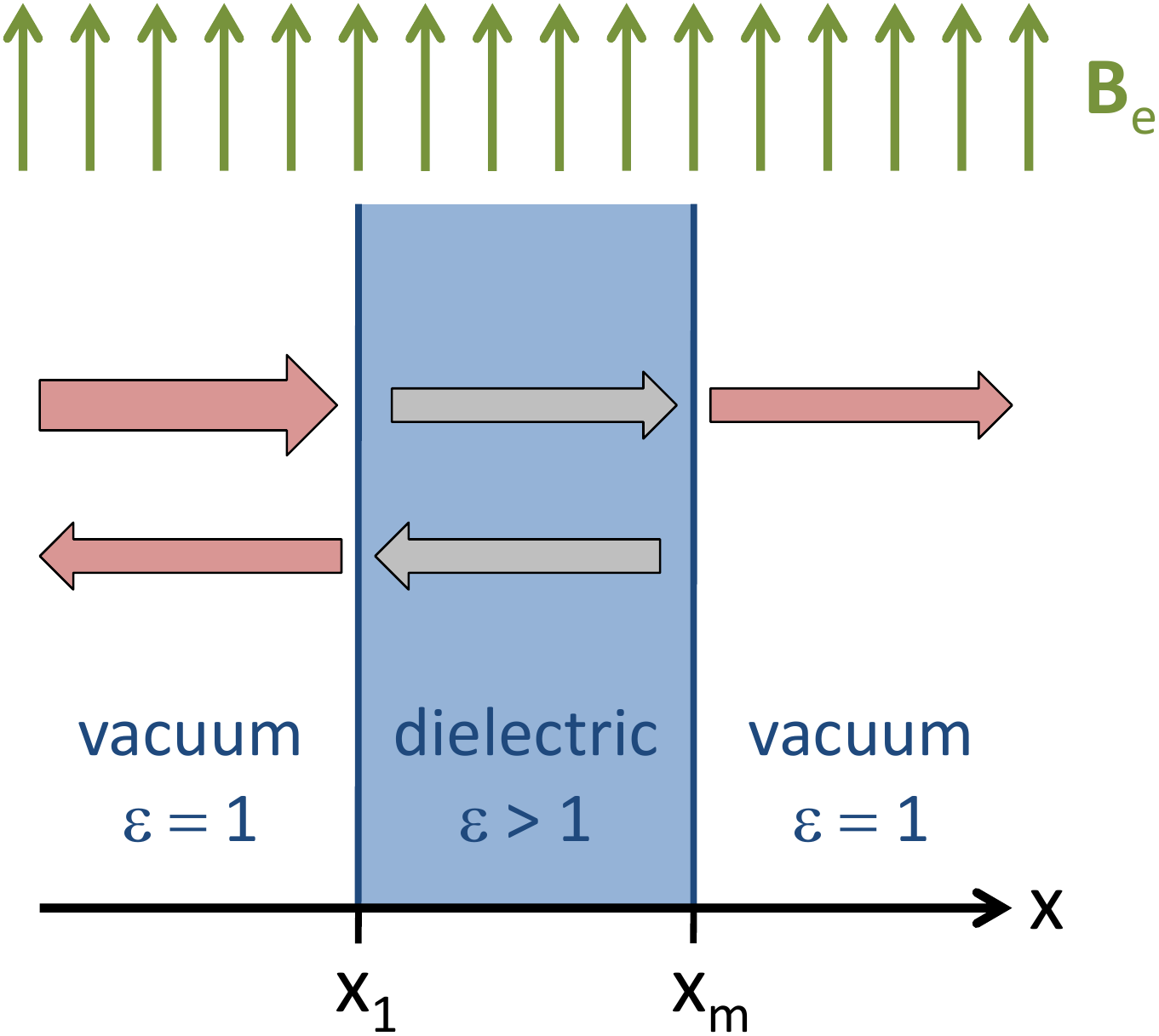}\hskip1.4cm
\includegraphics[width=6cm]{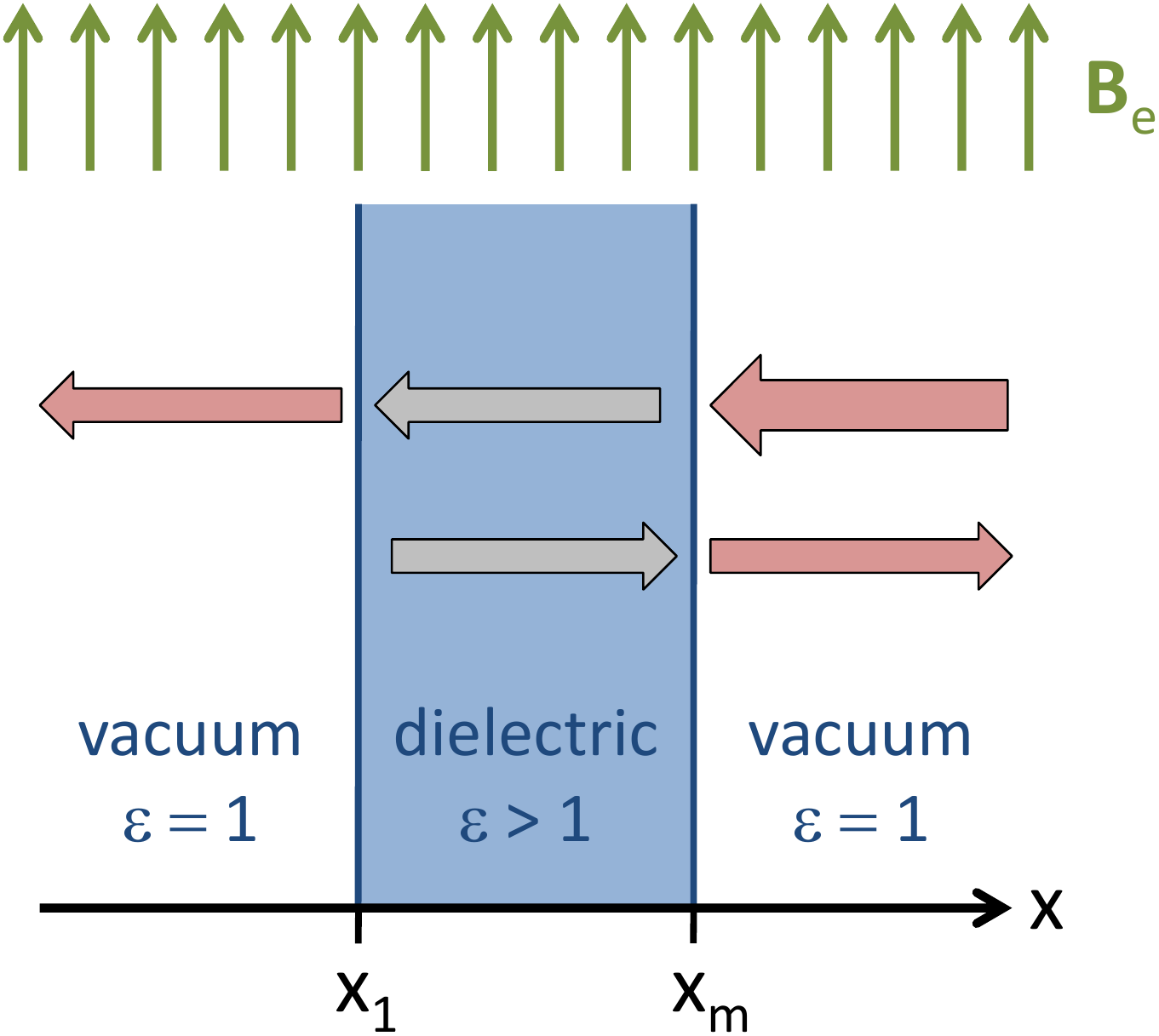}
\caption{Dielectric haloscope. The dielectric region could be a single layer as shown
or consist of many layers with the outermost interfaces at $x=x_1$ and $x=x_m$.
For several layers
there are many internal reflections which are not shown. The arrows
indicate the photon wave functions with an incoming wave being split into a
reflected and transmitted one. Reversing the arrows in both
panels, there exists a second set of wave functions 
with two incoming waves coalescing to a single outgoing one. The shown set
of wave functions 
or this second set are equivalent representations, as shown explicitly in reference~\cite{Garibyan:1961}.
}
\label{fig:disk}
\end{figure}

Our real interest, however, is a setup where the quasi-homogeneous
external magnetic field falls off adiabatically so that it does
not provide any significant momentum transfer and the
axion-photon transition rate vanishes with excellent approximation.
However, we now introduce a system of parallel dielectric layers,
oriented parallel to ${\bf B}_{\rm e}$. The simplest case of
a dielectric disk is shown in figure~\ref{fig:disk}. The dielectric
disk breaks translation invariance and thus enables the
transition between dark-matter axions and microwave photons.

It is now the photon wave function itself which is distorted by
the dielectric layers and thus no longer an eigenstate of
momentum. The same logic applies to transition radiation which
arises when a charged particle traverses dielectric interfaces.
While transition radiation is traditionally treated on the classical
level, a quantum calculation is required to treat backreaction on the
emitting particle. This point was first made by
Garibian~\cite{Garibyan:1961,Garibyan} who defined the photon wave functions
for the simple case of a dielectric interface, but the same
approach applies to a dielectric disk \cite{Grimus:1994ug} or more
complicated arrangements. We will only consider axions with
momentum ${\bf p}$ perpedicular to the interfaces, so they and the
photon wave functions will be translationally invariant in the
transverse directions.

\subsection{Garibian wave functions}

In this simple geometric setup, the Garibian wave functions consist of
an incoming plane wave which is split by the haloscope into a
transmitted and a reflected component. There are two such wave
functions, depending on the side of the haloscope on which they
impinge. Without the haloscope, these would be left- and right-moving
plane waves with wave vector ${\bf k}$. Equivalently, one can use the
time-reversed wave functions with all momenta reversed, corresponding
to two incoming waves which coalesce such as to produce only one
outgoing wave on either side of the haloscope. One can use one or the
other set to span the space of photon wave functions. The picture of
an incoming beam being split into two outgoing waves as shown in
figure~\ref{fig:disk} is somewhat more intuitive to construct the wave
functions, whereas the coalescence picture is more intuitive when we
ask for photon emission in one specific direction from the
haloscope. Notice that in general our system is not left-right
symmetric.

Explicitly we write the Garibian wave functions for the photon
electric field configuration for our one-dimensional problem in
the form ${\bf E}({\bf r})=i\omega{\bm \epsilon} A_\omega(x)$, where
${\bm \epsilon}$ is a polarization vector parallel to
${\bf B}_{\rm e}$. Outside of the haloscope
($x<x_1$ and $x>x_m$) the L wave function (left panel of figure~\ref{fig:disk})
has the form
\begin{equation}
  A^L_\omega(x)=\begin{cases}e^{i\omega \Delta x_1}+{\cal R}_L\,e^{-i\omega \Delta x_1}&\hbox{for $x<x_1$,}\\
  {\cal T}_L\,e^{i\omega \Delta x_m}&\hbox{for $x>x_m$,}
  \end{cases}
\end{equation}
corresponding to a plane wave with amplitude 1 impinging
from the left side and $\Delta x_j=x-x_j$. Notice that we define these coefficient  with reference to the left most interface ($x=x_1$) for the waves on the lhs of the device and rightmost interface ($x=x_m$) for those on the rhs of the device.
It is reflected with the reflection
coefficient ${\cal R}_L$ and transmitted with ${\cal T}_L$, where
$|{\cal R}_L|^2+|{\cal T}_L|^2=1$. In general,
our system is not left-right symmetric, so one needs to treat
separately the case where the incoming wave impinges from the right,
\begin{equation}
  A^R_\omega(x)=\begin{cases}{\cal T}_R\,e^{-i\omega \Delta x_1}&\hbox{for $x<x_1$,}\\
  e^{-i\omega \Delta x_m}+{\cal R}_R\,e^{i\omega \Delta x_m}&\hbox{for $x>x_m$.}
  \end{cases}
\end{equation}
In the haloscope region $x_1<x<x_m$, the wave functions can be very
complicated due to the multiple reflections on many layers.

The main feature of these wave functions is that the normalisation is such
that the incoming wave has amplitude 1, identical to an ordinary
plane-wave photon, independently of the detailed behavior within the
haloscope. This is physically clear if we think of a single photon
moving toward the haloscope and if we think of it as a wave packet
which does not yet know about the haloscope. It is built from ordinary
plane wave components which must be the same whether or not the
haloscope has been put in place. Therefore, the normalisation of its
Fourier components should not be affected by the presence of the haloscope.
By the same token, the Garibian wave functions for different $\omega$
and $\omega'$ must be orthogonal. The haloscope, being a linear optical element, cannot
mix photons with different frequencies.

The orthonormality of such wave functions was shown explicitly
for the case of a single dielectric disk for a more complicated situation
where the photons impinge at some angle
\cite{Grimus:1994ug}.  This more general case was needed to treat
transition radiation from a particle traversing the disk. Here we are considering 1D setups with many layers of dielectric material. While this situation appears more complicated, we show in Appendix~\ref{orthogonality} that one can use the relationship between the fields on either side of an interface to rewrite the orthonormality condition to be the same as for a free photon in a vacuum. This confirms that Garibian wave functions are the correct free photon wave functions for quantisation and use in our perturbative quantum calculation.

\subsection{Photon production rate}

To obtain the photon production rate we follow the same
steps as in section~\ref{sec:helioscope} for the helioscope, except
that the photon plane waves need to be replaced by Garibian wave
functions. In other words, the integral expression in
equation~(\ref{eq:matrix-helioscope}) must be substituted with
\begin{equation}
  {\cal I}_\omega=\int_{-\infty}^{+\infty} dx\,A_\omega(x)\,,
\end{equation}
where we have already assumed zero-velocity axions, i.e., ${\bf p}=0$,
implying $\omega=m_a$.

The integrals outside of the haloscope can be performed explicitly if we observe that
the oscillating part at infinity does not contribute, so for example
$\int_{x_m}^\infty dx\,e^{i\omega x}=i e^{i\omega x_m}/\omega$. Therefore the scattering amplitudes
are
\begin{subequations}\label{eq:amplitudes}
\begin{eqnarray}
  {\cal I}_\omega^L&=&\frac{1-{\cal R}_L-{\cal T}_L}{i\omega}
  +\int_{x_1}^{x_m} dx\,A_\omega^L(x)\,,\\[2ex]
  {\cal I}_\omega^R&=&\frac{1-{\cal R}_R-{\cal T}_R}{-i\omega}
  +\int_{x_1}^{x_m} dx\,A_\omega^R(x)\,.
\end{eqnarray}
\end{subequations}
In general, these expressions are different for L and R photons.

The photon flux per unit area and unit time emerging from the L or R
side of the haloscope therefore is
\begin{equation}
\Phi_{L,R}=\frac{\rho_a}{m_a}\,
\(\frac{g_{a\gamma}B_{\rm e}}{2}\)^2\,\left|{\cal I}_{\omega}^{L,R}\right|^2 \label{eq:overlapboost}
\end{equation}
with $\omega=m_a$.
In particular, the photon flux emerging from the L side is calculated
using the L photon as in the left panel of
figure~\ref{fig:disk}. Notice that we can look at this configuration as the
time-reversed case with a photon emerging on the L side, using these
as the appropriate ``out states.''

In order to connect to the previous literature we notice that in
the classical treatment of Ref.~\cite{Millar:2016cjp} the
photon flux was expressed in the form
\begin{equation}
\Phi_{L,R}=\frac{\rho_a}{m_a}\,
\(\frac{g_{a\gamma}B_{\rm e}}{m_a}\)^2\,\left|{\cal B}_{\omega}^{L,R}\right|^2\,,
\end{equation}
where ${\cal B}_{\omega}^{L,R}$ is the left or right ``boost amplitude''. It expresses
the electric field amplitude of the emerging EM wave in units of the
electric field induced by the galactic axion field within a magnetic field region.
Therefore, up to some global phase the connection to our scattering amplitude is
\begin{equation}
{\cal B}_\omega^{L,R}=\frac{\omega}{2}\,{\cal I}_\omega^{L,R}\,. \label{eq:boostfactor}
\end{equation}
The boost amplitude, like the transmission and reflection coefficients, is
a dimensionless quantity.

Thus we see that from a purely QFT starting point, we can obtain the same result of the classical analysis of reference~\cite{Millar:2016cjp}. While a similar expression to equation~\eqref{eq:overlapboost} was derived in reference~\cite{Millar:2016cjp}, it was not extended to small values of $|{\cal B}|$ as the integral terms at $\pm \infty$ were not handled correctly. Further, the connection between the integrand of ${\cal I}_\omega$ and the Garibian free photon wave functions was not made: classically the structure and normalisation of ${\cal I}_\omega$ could only be derived mathematically as a way to encode boundary conditions, without any physical underpinning.

\subsection{Perfect mirror}

For the simplest example, we can calculate the boost amplitude for a flat dish antenna, i.e., an interface between a mirror and vacuum~\cite{Horns:2012jf}. In this case, there are only two regions, and, for a perfect mirror, photon modes are only supported in the vacuum region. In this case only one of ${\cal I}_\omega^{L,R}$ is non-trivial and we do not have to worry about multiple interfaces. As the $E$-field must be zero at the mirror, ${\cal R}=-1$. Thus we very easily get that ${\cal B}=-i$. As the boost amplitude is defined so that a dish antenna has $|{\cal B}|=1$, we find a perfect agreement with reference~\cite{Millar:2016cjp}.

That both this overlap integral formalism and the transfer matrices used in reference~\cite{Millar:2016cjp} give the same result is not surprising. However, each formalism comes with very different physical interpretations. In the overlap integral formalism, the conversion of axions to photons occurs throughout the volume and axions and photons are in some sense treated separately. The free photon wave function satisfies Maxwell's equations by itself, for example canceling itself at the surface of the mirror.

The transfer matrix approach brings the interfaces to the forefront: the axion acts as a source of a discontinuity at each interface, requiring propagating EM waves to be emitted from the interface to satisfy Maxwell's equations. For a mirror, the axion induced $E$-field must cancel at the mirror with the propagating wave, giving $|{\cal B}|=1$. The volume only comes in via the distances between each interface, giving rise to interference effects (though power is generated throughout the device). This difference between the two pictures is highlighted in the present case of a mirror: a calculation that involves integrating over all space and one that only uses a single surface, with no volume or length scales, give the same result. Mathematically, this is explained by the fact that the volume integral can be rewritten as a sum over the surface terms coming from the boundaries of integration in each region~\cite{Millar:2016cjp}.

\subsection{Many layers}

To make our result yet more explicit we assume the haloscope to consist of $m-1$ dielectric
regions between $x_1$ and $x_m$, with parallel interfaces at $x_j$ with $j=1,\ldots,m$.
(We follow the convention of reference~\cite{Millar:2016cjp} where the region left of the
haloscope is region $j=0$ and the region to the right is $j=m$, here both taken to be
essentially vacuum with $n_0=n_m=1$.)
In each region, the refractive index is $n_j=\sqrt{\epsilon_j}$, assuming trivial
magnetic permeability $\mu_j=1$. In each region $j$ the wave function is written
in the form
\begin{equation}
A_j(x)=\alpha_j e^{i n_j\omega \Delta x_{j}}+\beta_j e^{-i n_j\omega \Delta x_{j}}\,, \label{eq:layerwavefunction}
\end{equation}
where $\alpha_j$ is the amplitude of the right-moving component and $\beta_j$ the left-moving one and we take $\Delta x_0=\Delta x_1$. We follow the same prescription as reference~\cite{Millar:2016cjp}, so that the field amplitudes $\alpha_j$ and $\beta_j$ of the right and left moving EM waves are defined at the left boundary of every region, except for $\alpha_0$ and $\beta_0$ which are defined at $x_1$, i.e., the leftmost interface. Therefore, apart from a polarization vector parallel to ${\bf B}_{\rm e}$, the
electric field of this EM wave is $E_j(x)=i\omega(\alpha_j e^{i n_j\omega \Delta x_j}+\beta_j e^{-i n_j\omega \Delta x_j})$, whereas the magnetic field, orthogonal to the electric one, is
$B_j(x)=i\omega\,n_j(e^{i n_j\omega \Delta x_j}-\beta_j e^{-i n_j\omega \Delta x_j})$.
At every interface, $E$ and $H$ parallel to the surface must be continuous, and because
in our case $H=B/\mu=B$, these conditions are explicitly
\begin{subequations}
\begin{eqnarray}
\alpha_{j-1}e^{i\omega n_{j-1}d_{j-1}}+\beta_{j-1}e^{-i\omega n_{j-1}d_{j-1} }&=&
\alpha_{j}+\beta_{j}\,,\\
n_{j-1}\(\alpha_{j-1}e^{i\omega n_{j-1}d_{j-1} }-\beta_{j-1}e^{-i\omega n_{j-1}d_{j-1}}\)&=&
n_{j}\(\alpha_{j}-\beta_{j}\)\,,
\end{eqnarray}\label{eq:transfer}
\end{subequations}
where $d_{j-1}=x_j-x_{j-1}$ is the thickness of each dielectric layer.
There are $m$ such pairs of equations for $m$ interfaces. For the case of L photons,
two additional equations
are $\alpha_0=1$ and $\beta_m=0$, whereas for $R$ photons we have $\alpha_0=0$ and
$\beta_m=1$, so overall we have as many equations as unknown amplitudes.

Note that for L photons the reflection coefficient is ${\cal R}_L=\beta_0^L$ and
the transmission coefficient ${\cal T}_L=\alpha_m^L$, with similar
expressions for R waves, so the scattering amplitudes of equation~(\ref{eq:amplitudes}) are found to be
\begin{subequations}
\begin{eqnarray}
  {\cal I}_\omega^L&=&\frac{1-\beta_0^L -\alpha_m^L }{i\omega}
 +\sum_{j=1}^{m-1}
  \frac{\alpha_j^L\(e^{in_j\omega d_{j}}-1\)
   -\beta_j^L\(e^{-in_j\omega d_{j}}-1\)}{i\omega n_j}\,,\\[2ex]
  {\cal I}_\omega^R&=&\frac{1-\alpha_m^R -\beta_0^R }{i\omega}
  +\sum_{j=1}^{m-1}
  \frac{\alpha_j^R\(e^{in_j\omega d_{j}}-1\)
   -\beta_j^R\(e^{-in_j\omega d_{j}}-1\)}{i\omega n_j}\,.
\end{eqnarray}
\end{subequations}
In general, these expressions are different for L and R photons.

\subsection{Single disk}

As a specific example we use a single dielectric disk of thickness $d$ as shown in figure~\ref{fig:disk},
i.e., two interfaces at $x_1=-d/2$ and $x_2=d/2$, For L photons we have $\alpha_0=1$ and
$\beta_m=0$ and otherwise find
\begin{subequations}
\begin{eqnarray}
\beta_0 &=&-\frac{(n^2-1)\sin\delta}{(n^2+1)\sin\delta+i\,2n\cos\delta}\,,\\[1ex]
\alpha_1&=&\frac{-2\,(n+1)}{(n+1)^2-(n-1)^2 e^{i 2\delta}}\,,\\[1ex]
\beta_1 &=&\frac{-2\,(n-1)\,e^{i\,2\delta}}{(n+1)^2-(n-1)^2 e^{i 2\delta}}\,,\\[1ex]
\alpha_2&=&\frac{i\,2n}{(n^2+1)\sin\delta+i\,2n\cos\delta}\,,
\end{eqnarray}
\end{subequations}
where $\delta=n\omega d$ is the phase accrued by a photon traversing the disk.
The transmission coefficient is ${\cal T}_L=\alpha_2$ and the reflection coefficient is
${\cal R}_L=\beta_0$.
With these explicit results it is straightforward to evaluate
the L scattering amplitude. We find for the corresponding boost factor
\begin{equation}
{\cal B}^L_\omega=-i\, \frac{(n^2-1)\,\sin(\delta/2)}{n^2\sin(\delta/2)+i\,n\cos(\delta/2)}\,,
\end{equation}
in agreement with reference~\cite{Millar:2016cjp} up to the same overall phase of $-i$ as in the dish antenna case.

\section{Quantum vs.\ classical calculation}
\label{sec:quantumclassical}

Thus far our calculation has assumed the transition from a single axion to a single photon,
whereas galactic dark matter axions have huge occupation numbers. If we consider an axion
plane wave with occupation number $N$, the axion part is
of the type $|N\rangle\to|N-1\rangle$, i.e., the quantized field amplitude part of the
matrix element is of the form $\langle N-1|a|N\rangle=\sqrt{N}$, where $a$ is the
destruction operator. Therefore, the transition rate now picks up a factor $N$, i.e.,
it is now proportional to $N/V$ with $V$ the normalisation volume.
In other words, we now have begun with $N$ axions instead of 1 and the photon
production rate remains proportional to the axion number density.

A given momentum mode of the axion field is highly occupied, but not necessarily in
a number eigenstate. The classical calculation uses a classical axion field that would
be represented by a Glauber state, i.e., a superposition of number states such that it
is an eigenstate of the destruction operator, not of the number operator, and the eigenvalue
would be the classical field strength in analogy to an optical laser.
Therefore, the photon
production rate is proportional to the square of the classical axion
field strength, a quantity that is the axion number density up to normalisation factors.

In general, the axion field is in some superposition of number states
of the type $\phi_{\bf p}^N|N,{\bf p}\rangle$ for all momentum modes,
where $\phi_{\bf p}^N$ is the amplitude for finding the
momentum mode ${\bf p}$ occupied with $N$ quanta. Because the
states $|N,{\bf p}\rangle$ are orthogonal, the overall rate is simply
proportional to $\sum_{N,{\bf p}}|\phi_{\bf p}^N|^2$ and thus again
to the total number of axions per normalisation volume.

So far we have assumed that the final-state photon state is empty before
production by axions, but hypothetically it could be occupied with
$N_\gamma$ quanta, and the rate would acquire a stimulation factor $N_\gamma+1$.
Actually there is always an ambient bath of thermal photons unless the
detector works at sufficiently cryogenic conditions.
However, in this case there is also a backconversion of photons to axions
which is proportional to $N_\gamma(N_a+1)$ with $N_a$ the axion occupation number.
So the net rate is proportional to $N_a(N_\gamma+1)-(N_a+1)N_\gamma=N_a-N_\gamma$.
In other words, the dominant part $N_aN_\gamma$ cancels and one cannot
achieve an increased stimulated conversion rate.

The bottom line is that, as long as we are only asking for the average
photon production rate, there is no difference between a classical calculation
using classical axion and photon fields and a quantum-field calculation using
quantized field amplitudes. A classical field has the boson property built in,
so no additional coherence or stimulation factors appear. On the other hand,
if one were not only to measure the average photon production rate but
also its fluctuation spectrum or time correlations, then the detailed
quantum properties of the axion field would feed through to the
secondary EM radiation.

\section{Conclusions}
\label{sec:conclusions}

We have calculated the microwave production rate in a dielectric haloscope
to search for axion dark matter. Our starting point was the ordinary perturbation
theory for any quantum transition rate. The key point was to use
distorted photon wave functions caused by the presence of dielectric interfaces.
In contrast to Primakoff axion-photon conversion or axion-photon oscillations
in large-scale magnetic fields where translational invariance is broken by
the external EM field and as such by the interaction term, it is here broken
by the photon wave function, regardless of the interaction term. This situation
is analogous to transition radiation at dielectric interfaces. We have used the
distorted wave functions first introduced by Garibian for a quantum calculation
of transition radiation. Our final expression for the production rate involves
an overlap integral between the distorted photon wave functions and
the magnetic field configuration, assuming a plane-wave initial axion.

The problem of microwave production in a dielectric haloscope was previously
addressed by solving the classical Maxwell equations of the coupled EM-axion system
in the presence of dielectric layers. Moreover, it was shown that the final
result could be transformed to an expression involving an overlap integral
between certain EM solutions of Maxwell's equations without axions
(see Appendix~B of reference~\cite{Millar:2016cjp}).
In other words, the Garibian wave functions were recovered somewhat by backward engineering
through cumbersome mathematical transformations.

Our results here provide a simple and direct physical underpinning for these findings and establish a direct connection to the analogous case of transition radiation. This allows us to understand many features of the overlap integral method that were somewhat mysterious in the classical treatment: questions such as the boundaries of integration, as well as the normalisation and choice of integrands, could only be derived mathematically, with no physical underpinning. This deeper physical understanding of the overlap integral formalism is made more interesting by the fact that the transfer matrices used in the classical treatment evoke a very different physical picture. Having a clear physical understanding of both pictures gives one greater flexibility to answer questions which may be obscured in one language.
Moreover, we have clarified that a classical calculation is equivalent to
a traditional first-order perturbative quantum calculation
so long as we only ask for the
average photon production rate.

\section*{Acknowledgments}

We thank Edoardo Vitagliano for helpful discussions. AI thanks the CERN theory department and the Werner Heisenberg Institute for their hospitality, where this work was done. We acknowledge partial support by the Deutsche Forschungsgemeinschaft
through Grant No.\ EXC 153 (Excellence Cluster ``Universe'')
and Grant No.\ SFB 1258
(Collaborative Research Center ``Neutrinos, Dark Matter, Messengers'')
as well as by the European Union through Grant No.\ H2020-MSCA-ITN-2015/674896
(Innovative Training Network ``Elusives'') and Grant No.\
H2020-MSCA-RISE-2015/690575 (Research and Innovation Staff Exchange project ``Invisibles Plus'').

\appendix
\section{Orthogonality of Garibian wave functions}
\label{orthogonality}
To check that the Garibian wave functions used in section~\ref{sec:haloscope} are the correct free photon wave functions, we must check that they are eigenvalues of the Hamiltonian with the appropriate energy. In particular, for our wave functions we must confirm that \cite{Grimus:1994ug}
\begin{equation}
\frac{1}{V\sqrt{\omega \omega '}}\int dV\frac{1}{2} \left[ \epsilon {\bf E}^*({\bf k}')\cdot{\bf E}({\bf k})+{\bf B}^*({\bf k}')\cdot{\bf B}({\bf k})\right]=\omega\delta_{{\bf k},{\bf k}'},\label{eq:orthogonality}
\end{equation}
where for simplicity we have assumed that $\mu=1$ throughout the device, and that $\epsilon$ does not depend on $\omega$. Here ${\bf k}$ denotes the wavenumber of the incoming/outgoing photon wave function (i.e., the wavenumber outside the device). Recall that throughout this paper we defined $\bf E$ and $\bf B$ essentially classically in terms of the unnormalised wave function $A_\omega$; the factor of $1/\sqrt{\omega V}$ must be restored to obtain a finite result from integrating over the Hamiltonian density. While this situation seems much more complicated then the case for plane waves, it will turn out that the contribution from every interface is exactly zero, allowing one to transform this integral into one over simple plane waves.

Consider arbitrary incoming waves of frequencies $\omega$ and $\omega'$, specified by some $\alpha_0',\beta_m'$ and $\alpha_0,\beta_m$. The primed coefficients correspond to waves of frequency $\omega'$ and the unprimed to waves of frequency $\omega$. The integrals in the $y,z$ directions are trivial, simply pulling out the area $S$. Using Eqn. \eqref{eq:layerwavefunction} we can write the remaining integral in a piecewise fashion,
\begin{eqnarray}
\frac{1}{2}\int dx\left [ \epsilon E^*(\omega') E(\omega)+B^*(\omega') B(\omega)\right ]&=&\omega \omega'\int_{-L/2}^{x_1}dx\, \alpha_0'^{*}\alpha_0 e^{i\Delta x_0\Delta\omega}+\beta_0'^{*}\beta_0 e^{-i\Delta x_0\Delta\omega}\nonumber \\
&&\kern-5em{}+\omega \omega'\sum_{j=1}^{m-1}n_j^2\int_{x_j}^{x_{j+1}}dx\, \alpha_j'^{*}\alpha_j e^{in_j\Delta x_j \Delta\omega}+\beta_j'^{*}\beta_j e^{-in_j\Delta x_j\Delta\omega}\nonumber \\
&&\kern-5em{}+\omega \omega'\int_{x_m}^{L/2}dx\, \alpha_m'^{*}\alpha_m e^{i\Delta x_m\Delta\omega}+\beta_m'^{*}\beta_m e^{-i\Delta x_m\Delta\omega}, \label{eq:fullint}
\end{eqnarray}
where $\Delta\omega=\omega-\omega'$.
Fortunately, it is sufficient to consider the evaluation of the integrals at the boundary and the interior integral associated with each interface separately. Ignoring the boundary terms for now, we can rewrite the interior integral as the sum over interfaces,
\begin{equation}
\frac{i\omega \omega'}{\Delta\omega}\sum_{j=1}^{m} n_{j}\left[\alpha_{j}'^{*}\alpha_{j} -\beta_{j}'^{*}\beta_{j} \right]- n_{j-1}\left[\alpha_{j-1}'^{*}\alpha_{j-1} e^{i d_{j-1} n_{j-1}\Delta\omega}-\beta_{j-1}'^{*}\beta_{j-1} e^{-id_{j-1} n_{j-1}\Delta\omega}\right], \label{eq:interiorintegral}
\end{equation}
where we have already performed the integration. The first two terms (with a factor of $n_{j}$) correspond to the integral being evaluated on the rhs of the interface and the last two terms (with a factor of $n_{j-1}$) correspond to evaluating the integral on the lhs. To see that all terms in this sum cancel, we can look more closely at the terms coming from the lhs of the interface. From equation~\eqref{eq:transfer}, we know that the $\alpha$ and $\beta$ on the lhs and rhs of an interface between media with different refractive indices are related. More specifically,
\begin{equation}
\vv{\alpha_j }{\beta_j }=\frac{1}{2n_{j}}\(\begin{array}{cc}
n_{j}+n_{j-1} & n_{j}-n_{j-1} \\
n_{j}-n_{j-1} & n_{j}+n_{j-1}  \end{array}\)\vv{\alpha_{j-1} e^{in_{j-1} d_{j-1}\omega}}{\beta_{j-1} e^{-in_{j-1} d_{j-1}\omega}}.
\end{equation}
With a little algebra, this allows us to rewrite the lhs interface terms using only quantities from the rhs of the interface,
\begin{eqnarray}
n_{j-1}\left[\alpha_{j-1}'^{*}\alpha_{j-1} e^{i d_{j-1} n_{j-1}\Delta\omega}-\beta_{j-1}'^{*}\beta_{j-1} e^{-id_{j-1} n_{j-1}\Delta\omega}\right] =n_{j}\left[\alpha_{j}'^{*}\alpha_{j} -\beta_{j}'^{*}\beta_{j} \right],
\end{eqnarray}
which makes it immediately clear that every term in equation~\eqref{eq:interiorintegral} vanishes. Thus the only terms contributing to the integral in equation~\eqref{eq:fullint} are the boundary terms evaluated at the edge of the normalisation volume ($\pm L/2$). To deal with these, we can use that conservation of energy requires the incoming and outgoing waves to be related by SU(2) matrices, and that we can switch left and right moving waves by changing the sign of the integration limits and $x$. This allows us to rewrite the integral in equation~\eqref{eq:fullint} as an integration over simple plane waves,
\begin{eqnarray}
\omega \omega'\int_{-L/2}^{L/2}dx\,\[\alpha_0'^{*}\alpha_0 e^{ix\Delta\omega}+\beta_m'^{*}\beta_m e^{-ix\Delta\omega}\]=\omega^2L(\alpha_0'^{*}\alpha_0+\beta_m'^{*}\beta_m )\delta_{\omega,\omega'},
\end{eqnarray}
where the equality holds for very large $L$ due to the oscillatory nature of the integral. Note that our Garibian wave functions are the only solutions that have definite asymptotic values of momentum, i.e., can be associated with a single traveling photon entering/exiting the haloscope. For these waves, as either $|\alpha_0|=1$ or $|\beta_m|=1$ so equation~\eqref{eq:orthogonality} must hold. Thus we see that our Garibian wave functions have the correct orthonormality conditions, and so can be used to quantise the photon field for use in section~\ref{sec:haloscope}.

\end{document}